%% file: main.tex
\documentclass[conference]{IEEEtran}
\IEEEoverridecommandlockouts

\RequirePackage[utf8]{inputenc}
\RequirePackage[english]{babel}
\RequirePackage{cite}
\RequirePackage{amsmath,amssymb,amsfonts}
\RequirePackage{algorithmic}
\RequirePackage[pscoord]{eso-pic}
\RequirePackage{textcomp}
\RequirePackage{xcolor}
\RequirePackage[a4paper, total={184mm,239mm}]{geometry}
\PassOptionsToPackage{detect-all, per-mode=symbol, range-units=single, range-phrase=~to~}{siunitx}
\RequirePackage{siunitx}
\RequirePackage{nicefrac}
\RequirePackage[hidelinks]{hyperref}
\RequirePackage{orcidlink}
\RequirePackage[noabbrev,capitalise]{cleveref}
\RequirePackage{float}
\RequirePackage[caption=false]{subfig}
\RequirePackage{url}
\RequirePackage{lipsum}
\RequirePackage{placeins}
\RequirePackage{xfrac}

\RequirePackage{amssymb}
\RequirePackage{pifont}

\RequirePackage{threeparttable}
\RequirePackage{booktabs}
\RequirePackage{colortbl}
\RequirePackage{tabularx}
\RequirePackage{makecell}
\RequirePackage{multirow}

\RequirePackage{graphicx}
\RequirePackage{glossaries}
\RequirePackage{xspace}

\glsdisablehyper{}

\newcommand{\ap}{\textsc{AXI-Pack}}%§

\newcommand{\x}{$\times$}

\renewcommand{\subsubsection}[1]{\paragraph*{\textbf{#1}}}

\DeclareSIUnit\GE{GE}
\DeclareSIUnit\kGE{\kilo\GE}
\DeclareSIUnit\MGE{\mega\GE}
\DeclareSIUnit{\bit}{b}

\begin{document}

% Acronyms
\input{acronyms.tex}

\newif\ifreviewmode

% Comment out to leave the review mode
% \reviewmodetrue

% Title
\title{Near-Memory Parallel Indexing and Coalescing: Enabling Highly Efficient Indirect Access for SpMV}

\ifreviewmode
    \author{%
        \vspace{0.05cm} %
        \textit{Authors omitted for blind review}
        \vspace{0.05cm} %
    }
\else
\author{
    \IEEEauthorblockN{%
    Chi Zhang\orcidlink{0000-0000-0000-0000}\textsuperscript{\textasteriskcentered}, % 
    Paul Scheffler\orcidlink{0000-0000-0000-0000}\textsuperscript{\textasteriskcentered}, % 
    Thomas Benz\orcidlink{0000-0002-0326-9676}\textsuperscript{\textasteriskcentered}, %
    Matteo Perotti\orcidlink{0000-0000-0000-0000}\textsuperscript{\textasteriskcentered}, %
    Luca Benini\orcidlink{0000-0001-8068-3806}\textsuperscript{\textasteriskcentered}\textsuperscript{\textdagger}%
    }
    \IEEEauthorblockA{
        \textasteriskcentered~\textit{Integrated Systems Laboratory, ETH Zurich}, Switzerland \\
        \textdagger~\textit{Department of Electrical, Electronic, and Information Engineering, University of Bologna}, Italy \\
        \{chizhang,paulsc,tbenz,mperotti,lbenini\}@iis.ee.ethz.ch
    }
}
\fi

\maketitle

\begin{abstract}
% required words count: <= 300 words for DATE
% current word count (update if changed): 198

% Do NOT use macros here; per IEEE guidelines, abstracts must be pure ASCII.
% Current Situation & Problem
Sparse matrix vector multiplication (SpMV) is central to numerous data-intensive applications, but requires streaming indirect memory accesses that severely degrade both processing and memory throughput in state-of-the-art architectures. Near-memory hardware units, decoupling indirect streams from processing elements, partially alleviate the bottleneck, but rely on low DRAM access granularity, which is highly inefficient for modern DRAM standards like HBM and LPDDR.
% Our solution
To fully address the end-to-end challenge, we propose a low-overhead data coalescer combined with a near-memory indirect streaming unit for AXI-Pack, an extension to the widespread AXI4 protocol packing narrow irregular stream elements onto wide memory buses. Our combined solution leverages the memory-level parallelism and coalescence of streaming indirect accesses in irregular applications like SpMV to maximize the performance and bandwidth efficiency attained on wide memory interfaces.
% Why are we better + Results
Our solution delivers an average speedup of 8x in effective indirect access, often reaching the full memory bandwidth. As a result, we achieve an average end-to-end speedup on SpMV of 3x. Moreover, our approach demonstrates remarkable on-chip efficiency, requiring merely 27kB of on-chip storage and a very compact implementation area of 0.2-0.3mm² in a 12nm node.
\end{abstract}

\begin{IEEEkeywords}
Sparse Computing, Memory Systems, Coalescing, DRAM, HBM, AXI4
\end{IEEEkeywords}

%
% Introduction
%
\section{Introduction}
\label{sec:intro}

\Gls{spmv} is pivotal in data-intensive application domains including machine learning\cite{Hoefler2021SparsityID}, fluid dynamics\cite{hpcg}, and graph analytics\cite{tdgraph}.
Sparse matrices in these domains often exhibit vast scales and low nonzero densities.
To reduce memory footprint, they are stored in compressed formats like \gls{csr} or \gls{sell}\cite{sellpack}.

%this introduces streaming indirect accesses in \gls{spmv}. 

Efficient \gls{spmv} poses a challenge to general-purpose architectures like \glspl{cpu}, \glspl{gpu}, and vector processors. 
% Efficient \gls{spmv} poses a challenge to a wide range of architectures, from general-purpose designs like \glspl{cpu}, \glspl{gpu}, and vector processors to dedicated \gls{spmv} accelerators. 
%
The challenge arises primarily from the indirect addressing and irregular, noncontiguous access of vector elements. Indirect addressing complicates program access flow as addresses depend on the values of indices in memory. Irregular memory access patterns disrupt the efficient utilization of the memory hierarchy in contemporary processors, resulting in low bandwidth efficiency, cache trashing, and elevated access latencies. 

Modern processors circumvent these issues by increasing core counts to saturate bandwidth, improving non-blocking cache miss handling, and increasing cache sizes. While these measures improve \gls{spmv} performance, they also incur significant additional on-chip area and further aggravate the cache pollution inherent to \gls{spmv}'s low data reuse\cite{spmv-cache-pollute}. 
% As a result, general-purpose \gls{spmv} compute utilization usually still remains below \SI{10}{\percent}\cite{}. 
%Even in the case of accelerators dedicated to \gls{spmv}, allocating vast on-chip memory (often exceeding \SI{10}{\mebi\byte} \cite{}) to store large, sparsely accessed vector chunks proves to be a suboptimal use of silicon area.

To address these inefficiencies, recent hardware proposals handle indirect accesses \emph{near memory}\cite{axi-pack,Carter1999ImpulseBA,Lloyd2015InMemoryDR,li2023accelerating}: as \gls{spmv}'s indirection can be fully decoupled from computation, indirected elements may be streamed directly from main memory without the need for large conventional caches. This insight has led to memory controllers with gather-scatter functionality~\cite{axi-pack,Carter1999ImpulseBA}, near-memory data layout transformers~\cite{Lloyd2015InMemoryDR}, and pattern-aware caches~\cite{li2023accelerating}. However, many of these solutions rely on \emph{narrow} memory channels with \emph{low access granularity} ($\leq$\,\SI{64}{\bit}) commonly provided by older \glsunset{dram}\gls{dram} standards, which enable efficient access to individual vector values in \gls{spmv}. 
In contrast, contemporary \gls{dram} interfaces usually have larger access granularities ($\sim$\,\SI{512}{\bit}); %
modern standards such as \gls{hbm} and \glsunset{lpddr}\gls{lpddr} do not provide narrow channels, instead prioritizing high bandwidth or low power. To achieve truly portable efficiency in \gls{spmv}, narrow elements must be efficiently accessed through the wide interfaces of these modern \gls{dram} subsystems.
%ignoring this inefficiency results in subpar bandwidth utilization and significant performance losses.

{\ap}~\cite{axi-pack} is a recently proposed extension to Arm's widespread \gls{axi}\cite{Arm2021AMBAAXI} protocol which maximizes \emph{on-chip} bus efficiency on \gls{spmv} and other irregular workloads. It packs multiple narrow elements onto a wide on-chip bus and enables \emph{bursts} of strided and indirect accesses, providing efficient on-chip transport of narrow data. Unfortunately, it was only demonstrated with low-granularity (\SI{32}{\bit}) on-chip scratchpads and does not address or accommodate the high granularity of modern off-chip \gls{dram}.

%proposed to support end-to-end bus-packed strided and indirect stream
%with full backward compatibility to \gls{axi}, but was designed to work only with multi-banked \glspl{sram} with a narrow (\SI{32}{\bit}) granularity.

In this work, we present an low-overhead access coalescer for {\ap} enabling highly efficient indirect accesses to modern, high-granularity \gls{dram} interfaces. To this end, we leverage the memory-level parallelism and coalescence of decoupled streaming indirect accesses, enabling cache-less data reuse and significantly improving bandwidth efficiency. With its aid, \gls{spmv} can be efficiently handled on general-purpose processors without the need for extensive on-chip resources.
% \todo{SELL ME the pen! To this end, we leverage the memory-level parallelism and coalescence of decoupled streaming indirect accesses, enabling cache-less data reuse and significantly improving bandwidth efficiency.}
%
%We want to retrieve indirect stream from main memory's high-grunarity interface in an on-chip efficient way.
Our contributions are as follows:
\begin{itemize}
    \item We extend the existing {\ap} adapter's indirect stream unit with a low-overhead data coalescer to accomodate a \SI{512}{\bit} granularity \gls{dram} controller. The new adapter design leverages both the memory-level parallelism and coalescence of indirect streams.
    \item Connecting our adapter to a standard \SI{512}{\bit} granularity \gls{hbm} channel controller, we achieve an 8\x~increase in effective indirect access bandwidth for real-world sparse matrices.
    \item We integrate our \gls{dram} coalescer into an open-source RISC-V vector processor system, where we achieve an average \gls{spmv} speedup of 10\x compared to a baseline with a \SI{1}{\mebi\byte} \gls{llc}, and 3\x over an adapter sans coalescer.
    \item Our solution shows 1.4\x and 2.6\x superior on-chip efficiency, while retaining 1\x and 0.9\x \gls{spmv} performance efficiency compared to state-of-the-art vector processors.
\end{itemize}

\begin{figure}[t!]
  \centering
  \includegraphics[width=1\linewidth, height=0.5\linewidth]{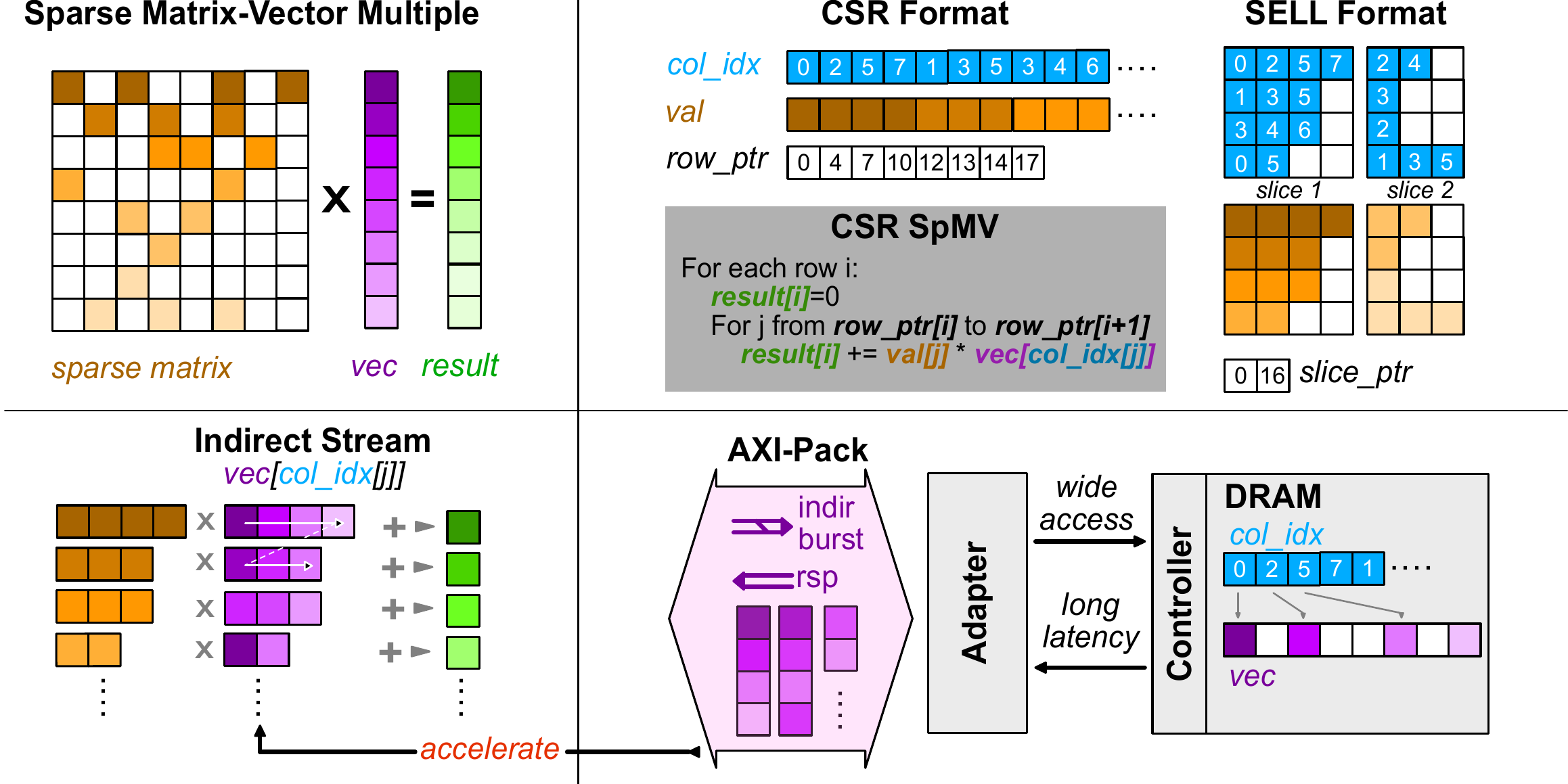}
  \caption{%
  (TL) shows a graphical representation of \gls{spmv}, a sparse matrix multiplied by a dense vector equals a dense vector. %
  (TR) represents the two file formats used in this work: \gls{csr} and \gls{sell}. %
  (BL) is the indirect stream of CSR \gls{spmv} accelerated by (BR) our coalescing \emph{adapter} using {\ap}. % 
  }
  \label{fig:spmv_bg}
\end{figure}

%
% Architecture
%
\section{Architecture}
\label{sec:arch}

\begin{figure*}[t!]
    \centering
    \subfloat[\label{fig:arch:a}]{{\includegraphics[height=5.0cm]{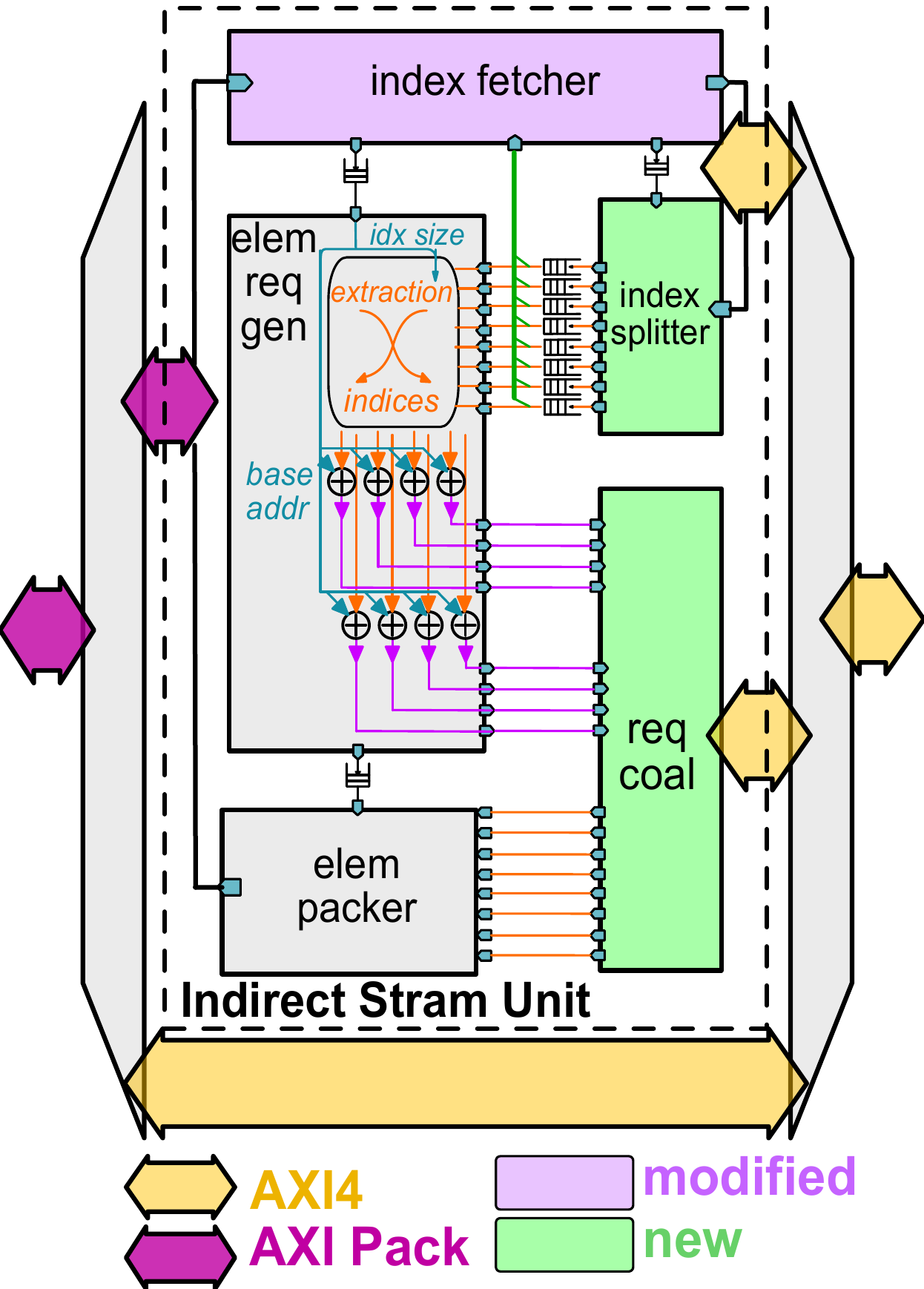}}}%
    \quad
    \subfloat[\label{fig:arch:b}]{{\includegraphics[height=5.0cm]{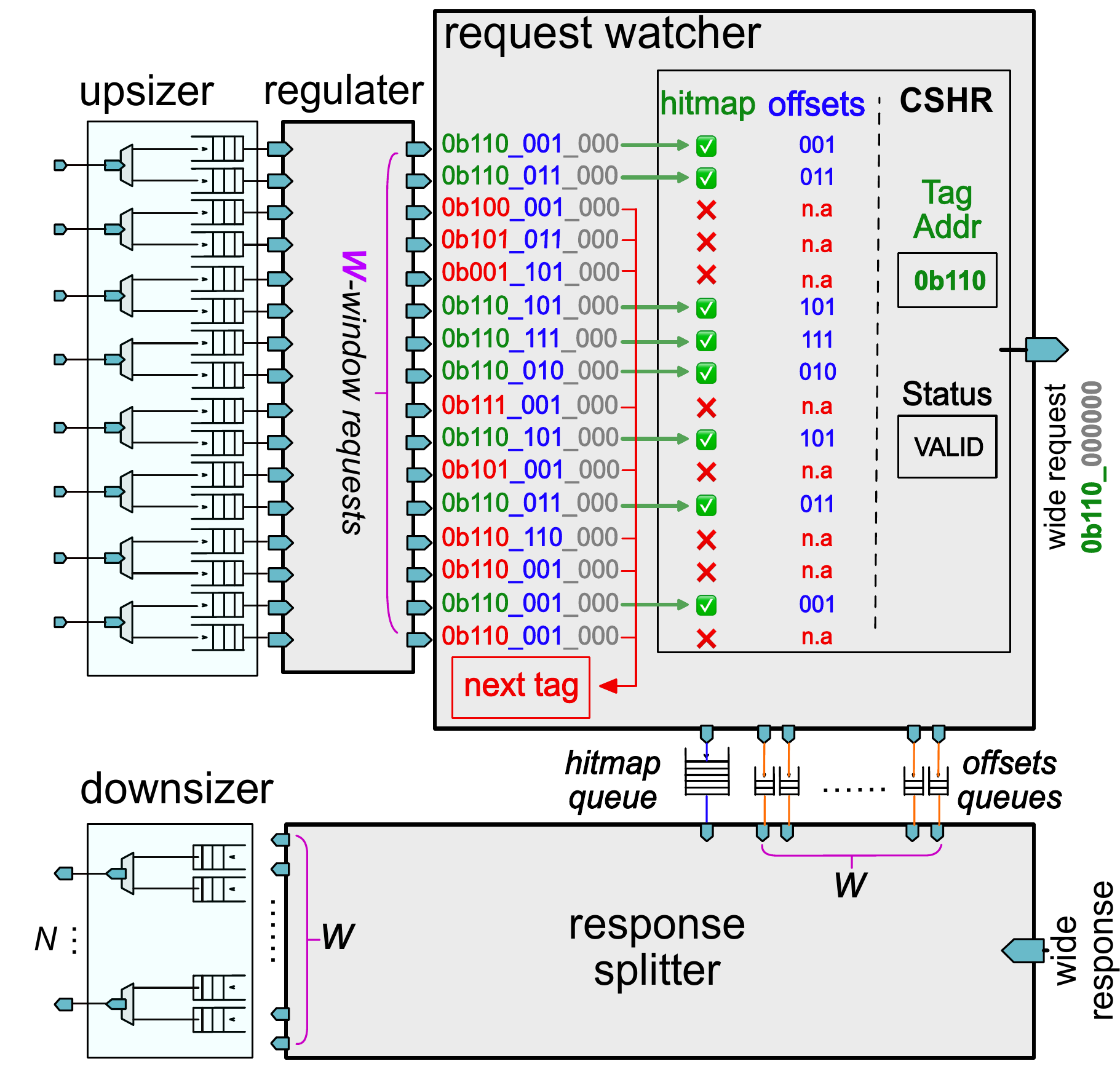}}}%
    \quad
    \subfloat[\label{fig:arch:c}]{{\includegraphics[height=5.0cm]{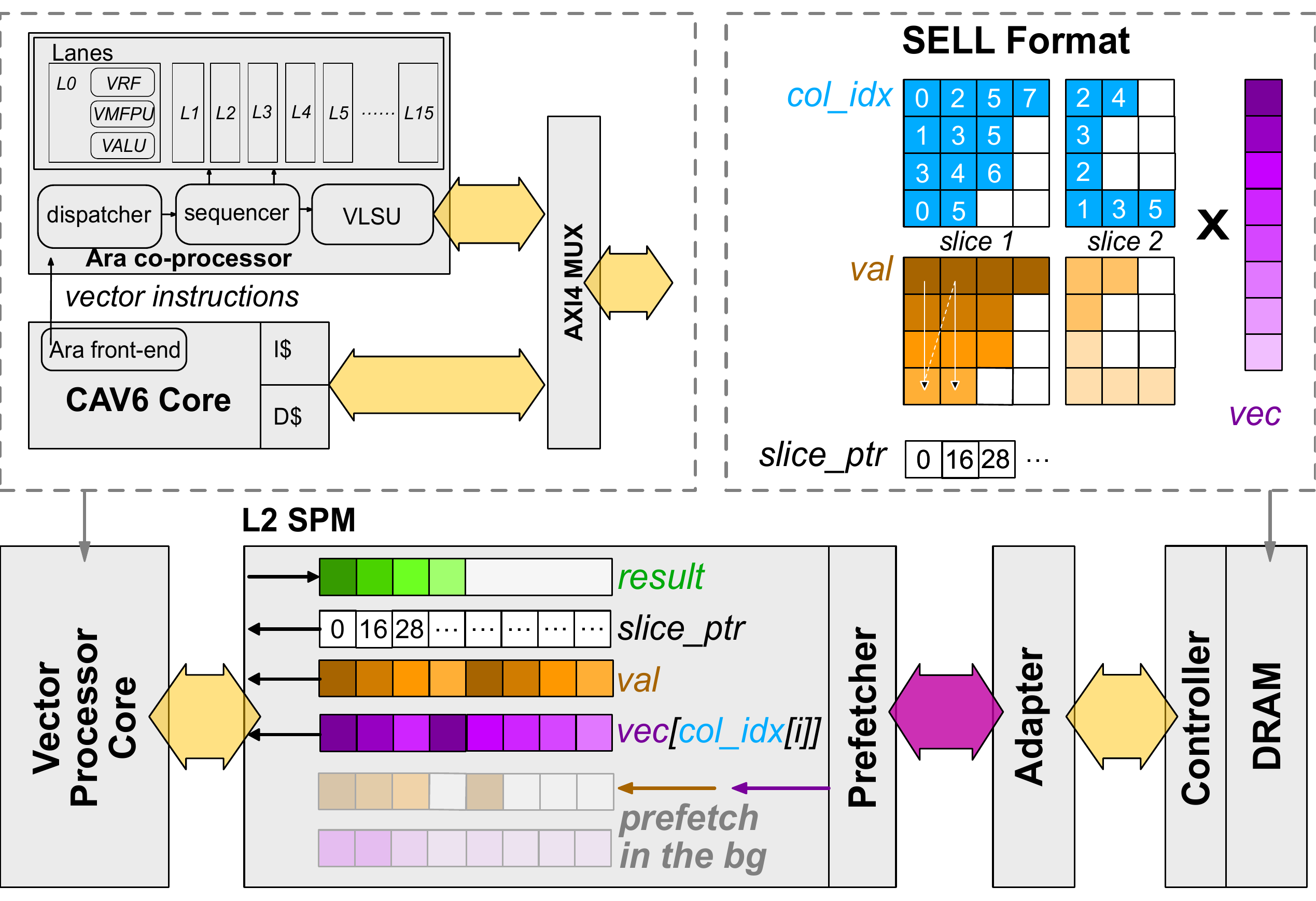}}}%
    \caption{%
        (a) shows the extended {\ap} adapter unit; modified blocks are marked in green and light purple. %
        (b) presents a detailed architectural view of the \emph{request coalescer (req coal)}. %
        (c) outlines the integration of our extended {\ap} in a system based around \emph{CVA6}~\cite{cva6} and \emph{ARA}~\cite{new_ara}. %
    }
    \label{fig:arch}
\end{figure*}

We will first present our \emph{indirect stream unit} translating {\ap} indirect bursts into wide \gls{dram} requests in \Cref{subsec:isu}. We will then detail the contained \emph{request coalescer} in \Cref{subsec:req_coal}. Finally, we will describe the \emph{integration} of our hardware into a vector processor system in \Cref{subsec:vpc}.

\subsection{{\ap} Indirect Stream Unit}
\label{subsec:isu}

\Cref{fig:arch:a} depicts the {\ap} adapter containing our \emph{indirect stream unit}, which is based on the previously proposed {\ap} indirect read converter \cite{axi-pack}. It translates {\ap} indirect burst requests into bandwidth-efficient sequences of wide \gls{dram} accesses in a non-blocking, index-parallel manner.
Among its five components, 
%are connected through decoupling queues as shown. 
%
the \emph{element request generator} and \emph{element packer} are reused from the original indirect read converter without modification. 
The \emph{index fetcher} was modified and the \emph{index splitter} and \emph{request coalescer} were newly added to efficiently interface wide \gls{dram} controllers downstream.

Upon receiving an indirect burst request from an upstream {\ap} manager, the \emph{index fetcher} determines the index stream's start address and length to  issue efficient wide \gls{dram} reads retreiving the indices. It monitors the usage of downstream index queues to prevent their overflow.
For every received wide block of indices, the \emph{index splitter} splits it into $N$ parallel segments, which are subsequently pushed into the aforementioned $N$ parallel index queues.
The \emph{element request generator} then produces $N$ parallel narrow element requests by extracting indices from the index queues and adding them to the requested element base address.

The generated $N$ parallel narrow element requests are passed to the \emph{request coalescer} which forms wide \gls{dram} accesses, reducing access redundancy and enabling data reuse in the process. We explain the inner workings of the coalescer in \Cref{subsec:req_coal}. Ultimately, it returns the $N$ parallel requested elements in the order of the original narrow requests.
To comply with the {\ap} protocol, the retrieved narrow elements are densely packed onto the upstream wide bus by the \emph{element packer}, maximizing on-chip bus utilization.

\subsection{Request Coalescer}
\label{subsec:req_coal}

\Cref{fig:arch:b} shows the architecture of the request coalescer. 
The \emph{upsizer} receives and pushes the $N$ incoming parallel narrow requests into $W$ decoupled request queues. 
Both $N$ and $W$ must be powers of two and $W{\geq}N$. 
For each narrow request port, requests are evenly distributed among $W/N$ request queues, collecting \emph{windows} of $W$ requests.
A \emph{regulater} presents a complete window of the $W$ oldest requests to the \emph{request watcher}. In case there aren't enough requests to form a complete window within a time limit, it forwards the remaining requests.

The \emph{request watcher} identifies requests in the current window accessing the same wide \gls{dram} block, together forming a request \emph{warp}. 
A \emph{\gls{cshr}} stores the following information on the current request warp:
\begin{itemize}
    \item \textbf{Tag}: 
    The address of the wide block in \gls{dram} for which requests are currently being coalesced. 
    If a narrow request in the current window can be served from this block, we call it a \emph{hit} request; otherwise, it is a \emph{miss} request.
    
    \item \textbf{Status}: 
    0 (\emph{IDLE}) denotes that a coalescing is in progress and 1 (\emph{VALID}) that a block has been coalesced.
    
    \item \textbf{Hitmap}: 
    A \emph{W}-bit array denoting which window requests have been merged into the current request warp.
    
    \item \textbf{Offsets}: 
    The address offsets for each coalesced request.
\end{itemize}

The request watcher only contains one active \gls{cshr}. 
In each cycle, the request watcher checks all valid requests in the window in parallel to see if they can hit the current \gls{cshr}. 
It then accepts hit requests, updates the \gls{cshr}'s \emph{Status}, \emph{Hitmap}, and \emph{Offsets} information accordingly, and tells the regulator to invalidate the corresponding window entries. 
With pending miss requests, the request watcher issues the wide request of the current \gls{cshr} downstream, updating the \emph{Hitmap} and \emph{Offsets} information to \emph{meta data queues}. 
Meanwhile, miss requests are used to determine the next \gls{cshr}'s \emph{Tag Address}.
Once all requests in the window have been coalesced, the \gls{cshr} tells the regulator to forward the next window. 
Additionally, a watchdog will issue the current \gls{cshr}'s wide request should no new request window arrive within a configurable time limit.

The \emph{meta data queues} consist of a deep \gls{fifo} holding the \emph{Hitmap} and \emph{W} shallow \glspl{fifo} for the \emph{Offsets}. 
Their behaviors differ slightly: every time the request watcher issues a wide access, it loads the entire \emph{Hitmap} array into the deep queue while only pushing \emph{offsets} of the hit window entries to the corresponding shallow \glspl{fifo}. 

In the return path, a response splitter retrieves the parallel narrow elements from the wide downstream data response. 
Upon receiving a response, it checks the \emph{Hitmap} queue to identify the window entries receiving the element. 
Subsequently, it pops the related \emph{Offsets} queues to retrieve the word offsets. 
It then extracts the elements of the window entries in parallel and forwards them to the \emph{W} \emph{element queues}. 
Finally, a downsizer mapps the  \emph{W} \emph{element queues} onto the \emph{N} output ports. 
To summarize,  the downsized is structurally similar to the upsizer performing the inverse function while enforcing the ordering.

\subsection{Integration into a Vector Processing System}
\label{subsec:vpc}

To evaluate its performance benefits, we integrate our {\ap} adapter into a {RISC-V} vector processor system as shown in \Cref{fig:arch:c}.
{Ara}~\cite{new_ara} is an open-source vector coprocessor supporting version \emph{1.0} of the {RISC-V} vector extension.
Together with the \SI{64}{\bit} Linux-capable {CVA6}~\cite{cva6} core, it forms a \gls{vpc} wherein {CVA6} offloads vector instructions to {Ara}. 
Both share a common \gls{axi}-based memory interface. 

The \gls{vpc}'s \SI{512}{\bit}-wide \gls{axi} interface is connected through a prefetching-enabled \gls{l2} \gls{spm} to \gls{dram} main memory.
The \gls{l2}'s \emph{prefetcher}'s manager interface implements a \SI{512}{\bit}-wide {\ap} bus connected to main memory through our coalescing-enhanced {\ap} \emph{Adapter}.  
This \emph{prefetcher} implements both contiguous and indirect prefetching through {\ap} requests. 
This unit is configured through {CVA6} and supports up to two outstanding prefetching requests.

%Using the \gls{sell} format, we demonstrate tiled \gls{spmv} execution in \Cref{fig:arch}.
We execute tiled \gls{spmv} workloads on the \gls{vpc} on data stores in \gls{dram} using the \gls{sell} format.
Six equally-sized arrays are allocated in the \gls{l2} \gls{spm}: one for the \emph{slice pointers} and the \emph{results}, two for holding the \emph{non-zero elements} and two for the \emph{indexed vectors}. 
Initially, the first data tile, consisting of the slice pointers, non-zero elements, and indexed vectors, is prefetched into \gls{l2} through {\ap} interface. 
The \gls{vpc} then prefetches the subsequent data tiles during the computation of the first tile.
Thanks to the indirect accesses of {\ap}, the \gls{vpc} mainly executes \gls{vmac} operations. 
%As the vector elements are already indexed by the {\ap}, the kernel mainly executes \gls{vmac} tasks. 
Slice pointers are consumed more slowly; thus, it is not required to prefetch them during the \gls{vpc}'s computation in every iteration.
The \gls{vpc} interrupts execution should the slice pointer array deplete or if the result array is full. %
The \gls{vpc} then signals the prefetcher to refresh the stored data in \gls{l2} \gls{spm}.

%
% Results: Methodology
%
\section{Evaluation Methodology}
\label{sec:eval}

\begin{table}[h!]
\centering
\caption{Adapter and Vector Processor System Parameters}
\label{tab:my_table}
\begin{tabularx}{\linewidth}{l|X}
\toprule
\textbf{Model} & \textbf{Parameter} \\
\midrule
\multirow{3}{*}{{\ap} Adapter} & Queue depth = 256(\emph{index}), 2(\emph{up/downsizer}), \\
& 128(\emph{hitmap}), 2048/\emph{W}(\emph{offsets}) \\
& On-chip storage = 27KB (\emph{W}=256) \\
\midrule
Vector Processor System & 16 lanes, 1GHz, 384KB L2\\
\midrule
\multirow{2}{*}{DRAM and Controller}  & One HBM2 chan, 1GHz, 32GB/s (ideal) \\
& Schedule policy: open adaptive, FR-FCFS\cite{dramsys} \\
\bottomrule
\end{tabularx}
\end{table}

% \begin{table}[h!]
% \centering
% \caption{Adapter and Vector Processor System Parameters}
% \label{tab:my_table}
% \begin{tabularx}{\linewidth}{l|X}
% \toprule
% \textbf{Parameter} & \textbf{Value} \\
% \midrule
% \multicolumn{2}{l}{\textbf{Adapter Parameters}} \\
% \midrule
% \multirow{5}{*}{Queue depth} & index: 256 \\
% & upsizer:   2 \\
% & downsizer: 2 \\
% & hitmap:    128 \\
% & offsets:   2048/\emph{w} \\
% On-chip storage (\emph{w}=256) & 27KB \\
% \midrule
% \multicolumn{2}{l}{\textbf{Vector Processor System}} \\
% \midrule
% Frequency & 1GHz \\
% Number of lanes in Ara & 16 \\
% L2 SPM size & 384KB \\
% DRAM model & one HBM2 chan \\
% & 1GHz, 32GB/s (ideal) \\
% DRAM CTRL page policy & Open Adaptive \\
% DRAM CTRL scheduling policy & first ready first come first serve(FR-FCFS) \\
% \bottomrule
% \end{tabularx}
% \end{table}

We built our {\ap} adapter and vector processor system using \gls{rtl} models, except for the \gls{dram} controller and models, which are modeled by \emph{DRAMSys}~\cite{dramsys}. We compiled DRAMSys into a dynamically linkable library and used {Siemens' Questa advanced simulator} to connect it, enabling co-simulation of our \gls{rtl} model with \gls{dram} models in a cycle-accurate manner. All our model parameters are listed in the \Cref{tab:my_table}. The adapter also has three variant types for comparison: without a coalescer (\emph{MLPnc}), with an \emph{x-window} parallel coalescer (\emph{MLPx}), and with an \emph{x-window} coalescer but operating sequentially (\emph{SEQx}). The latter configuration is achieved by serializing the parallel element requests and reducing the input ports of the coalescer to one. 
For indirect stream and \gls{spmv} benchmarking, we selected twenty real-world sparse matrices from the \emph{SuiteSparse}~\cite{sparsesuit} collection and \emph{HPCG} benchmark\cite{hpcg}, with columns ranging from 1.4k to 6.8M and nonzeros from 23k to 37M. When converting these matrices to CSR or SELL formats, we utilized \SI{32}{\bit} indices and \SI{64}{\bit} for nonzeros and metadata, with 32 rows per slice in SELL format. We also introduced a baseline for \gls{spmv} comparison on vector processor systems. This baseline uses a \SI{1}{\mebi\byte} \gls{llc} between the vector processor and memory controller, both connected by a 512\si{\bit}-wide \gls{axi} bus. Since the baseline lacks an indirect prefetching function, it executes the naive \gls{spmv} code with coupled indirect access and arithmetic operation.

%
% Results
%
\section{Experimental Results}
\label{sec:res}

\subsection{Indirect Stream Analysis}
\label{subsec:indr_ana}

\begin{figure*}[t!]
  \centering
  \includegraphics[width=1\linewidth, height=0.3\linewidth]{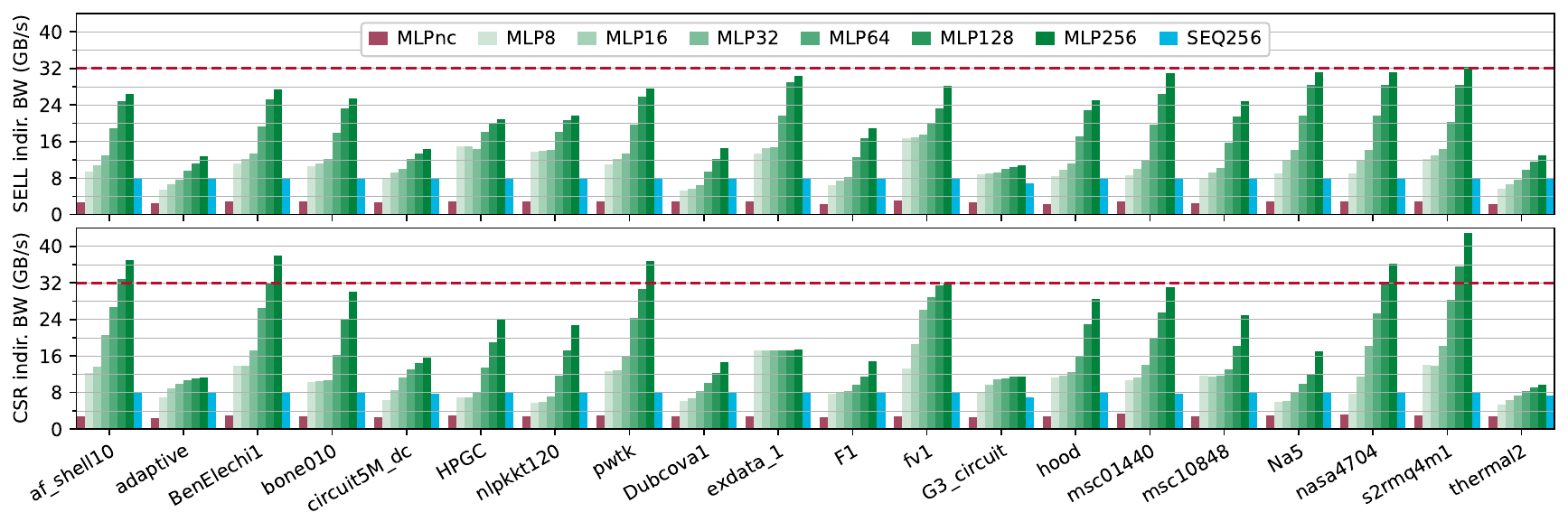}
  \caption{indirect stream bandwidth}
  \label{fig:isb}
\end{figure*}

\begin{figure}[t!]
  \centering
  \includegraphics[width=1\linewidth]{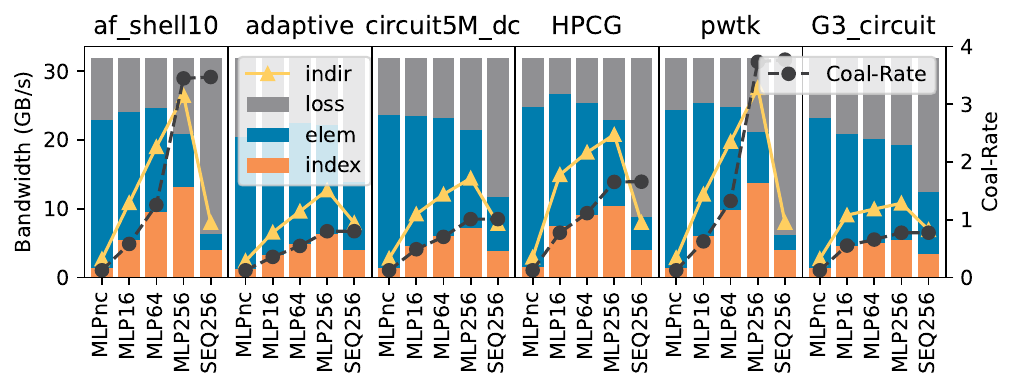}
  \caption{bandwidth breakdown and coalesce rate}
  \label{fig:bb}
\end{figure}

We first evaluate our adapter's ability to stream indirect accesses for whole \gls{spmv} dataflow. For testing, an ideal requestor issued continuous {\ap} indirect read requests from upstream, and our matrices, prepared in either \gls{sell} or \gls{csr} format, were preloaded into the \gls{hbm} model. 

As depicted in \Cref{fig:isb}, without coalescence, our design's indirect stream bandwidth averaged just \SI{2.9}{\giga\byte\per\second} out of the possible \SI{32}{\giga\byte\per\second}, a limitation stemming from the under-utilization of wide accesses. However, introducing coalescence drastically boosted this bandwidth, especially as the coalescence window size increased. Impressively, a 256-window parallel coalescer amplified the indirect stream bandwidth by factors of 8.4\x~and 8.6\x~for \gls{sell} and \gls{csr} formats, respectively. This enhancement enabled twelve of the twenty matrices to achieve an indirect access bandwidth surpassing \SI{70}{\%} of the full memory bandwidth. Some \gls{csr} formats even exceeded this limit, showcasing the 256-window parallel coalescer's prowess in harnessing data reuse. In contrast, a 256-window sequential coalescer cannot enhance performance as significantly as the parallel coalescer. Although it can still speed up the indirect stream by an average of 2.9\x  compared to a non-coalescer design, the indirect stream bandwidth is capped under \SI{8}{\giga\byte\per\second}, averaging a 3.0\x  slowdown compared to a parallel coalescer with the same window size.

To further comprehend the interplay of memory-level parallelism and coalescence, we delved deeper with six representative matrices (\gls{sell} format). Our focus was on the downstream bandwidth utilization and the adapter's coalesce rate. The bandwidth was categorized into three segments: fetching elements, fetching indices, and the loss from the ideal \gls{hbm} channel bandwidth. We defined the coalesce rate as the ratio of effective indirect access elements to the data amount requested by the coalescer from downstream. 

%In cases where element requests are contiguous, the coalesce rate is one, since all wide accesses for element fetching are fully utilized. Interestingly, our coalescer could also detect data reuses within its window, exemplified when two indices target the same element, pushing the coalesce rate above one in some instances.

\Cref{fig:bb} sheds light on several nuances. Without a coalescer, each narrow element request led to one wide access, monopolizing a large chunk of the downstream bandwidth for element fetching. This dominance compromised bandwidth reserved for fetching indices, while the effective indirect stream bandwidth remains low.
%The coalesce rate in this setup was a mere \emph{0.125}, indicating that only \SI{64}{\bit} data was gleaned from every \SI{512}{\bit} wide access. Despite downstream request saturation, overheads like page misses and refreshing in \gls{hbm} meant that only \SI{72.5}{\%} of the \gls{hbm} bandwidth was utilized.
By contrast, the parallel coalescer's deeper window boosted the coalesce rate. It coalesced more narrow requests under a larger window size while simultaneously reducing wide accesses for elements. This in turn allocated more downstream bandwidth for index fetching, boosting overall throughput. For instance, the matrix \emph{af-shell10} with the 256-window parallel coalescer, showcased an indices fetching bandwidth of \SI{13.2}{\giga\byte\per\second}. This suggests our adapter was concurrently generating and coalescing \emph{3.3} requests per cycle. Conversely, while the sequential 256-window coalescer might reach the same coalesce rate, the sequential processing becomes a significant bottleneck. One request per cycle limits the indices fetching bandwidth to around \SI{4}{\giga\byte\per\second} and reduces the bandwidth for element fetching, leading to a relatively low \gls{hbm} bandwidth utilization.

From our observations, memory-level parallelism and coalescence emerge as complementary forces for indirect stream:
\begin{enumerate}
\item Without coalescence, the bandwidth dedicated to index fetching is overshadowed by that for element fetching. Thus, having the hardware to generate parallel element requests from indices becomes redundant.
\item Absent memory-level parallelism, a sequential process can't capitalize on coalescence's potential, culminating in limited bandwidth utilization and reduced effective indirect access bandwidth.
\end{enumerate}

In conclusion, to maximize indirect access bandwidth, a symbiotic approach, harnessing both memory-level parallelism and coalescence, is indispensable.

\subsection{SpMV Performance}
\label{subsec:spmv_perf}

\begin{figure}[t!]
\centering
    
  \subfloat[Ara \gls{spmv} performance]{
    \label{fig:spmv-perf}
    \includegraphics[width=1\linewidth, height=0.325\linewidth]{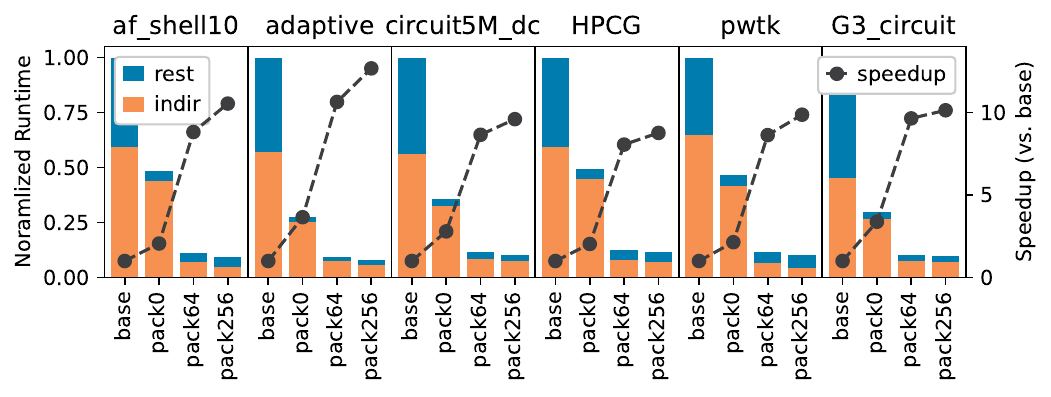}
  }
  
  \subfloat[\gls{spmv} off-chip traffic and bandwidth utilization]{
    \label{fig:octbu}
    \includegraphics[width=1\linewidth]{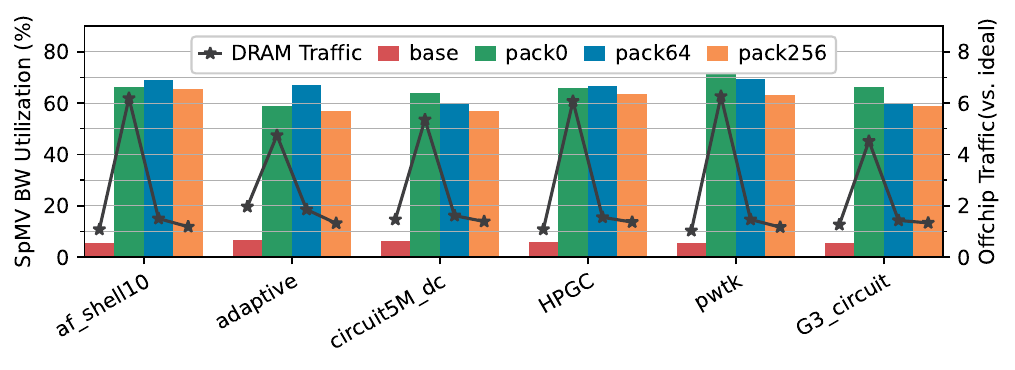}
  }

  \caption{\gls{spmv} characterization}
  \label{fig:spmv_character}
\end{figure}

% \begin{figure}[t!]
%   \centering
%   \includegraphics[width=1\linewidth]{fig/spmv_perf.pdf}
%   \caption{Ara \gls{spmv} performance}
%   \label{fig:spmv-perf}
% \end{figure}

% \begin{figure}[t!]
%   \centering
%   \includegraphics[width=1\linewidth]{fig/res_spmv_bwoc.pdf}
%   \caption{\gls{spmv} off-chip traffic and bandwidth utilization}
%   \label{fig:octbu}
% \end{figure}

To assess the impact of our adapter on \gls{spmv} performance, we benchmarked \gls{spmv} on four {RISC-V} vector processor systems. These include three {\ap} enabled vector processor systems (as described in \cref{subsec:vpc}) with configurations of no coalescer, and parallel coalescers with windows of 64 and 256 (labeled \emph{pack0}, \emph{pack64}, \emph{pack256}). Additionally, we used one \emph{base} system detailed in the \cref{sec:eval}.

In \Cref{fig:spmv-perf}, we show their \gls{spmv} speedup compared to the \emph{base}, normalized \gls{spmv} runtime, and the time spent on indirect accesses. The latter is counted from the prefetcher as the time to transfer the indirect stream in the \emph{pack} system. However, for the \emph{base} system, since it doesn't have a prefetcher, we count the indirect access time as the time the Vector-Load-Store-Unit spends on indices fetching and the gather operation.

The \emph{pack0} system demonstrates an average 2.7\x speedup compared to the \emph{base} system. We observed that even a naive prefetching scheme is better than using a conventional cache for \gls{spmv} to hide the access latency from \gls{dram}. It saves a significant amount of runtime in tasks like accessing non-zero elements, where the cache suffers from its lack of data reuse nature while the prefetcher leverages long data stream transfers in the background of \gls{spmv} arithmetic computation. However, it doesn't significantly reduce the time for indirect accesses since its {\ap} adapter isn't equipped with a coalescer, and the indirect stream bandwidth remains low.

In the \emph{pack64} and \emph{pack256} systems, we noticed a substantial improvement in \gls{spmv}. The increase of the coalescer window significantly shrinks the time for indirect access, which in turn reduces the runtime of \gls{spmv}. The \emph{pack256} system shows an average 3\x speedup compared to the \emph{pack0} system and a 10\x speedup compared to the \emph{base} system. 

%In the \emph{pack256} system, the time for indirect access is almost the same as the rest of the runtime, suggesting that having a coalescer with a window larger than \emph{256} is not necessary due to the diminishing effect of reducing indirect access time.

\Cref{fig:octbu} displays \gls{spmv}'s off-chip traffic and \gls{hbm} bandwidth utilization for all these systems compared to the ideal, which is no redundant off-chip traffic and full memory utilization. Although the \emph{base} system demonstrated low off-chip traffic overhead due to a large \gls{llc}, its memory utilization, as low as \SI{5.9}{\%}, indicates its poor performance. While the \emph{pack0} system has the best average memory utilization of \SI{65.8}{\%}, it exhibits an average 5.6~\x off-chip traffic compared to the ideal case, significantly increasing the energy waste on off-chip data movement while the performance is suboptimal. With the help of increasing the coalescer window size to 256, we largely reduced the off-chip traffic to only \SI{29}{\%} more than the ideal, while maintaining an average memory utilization of \SI{61}{\%}. Compared to the \emph{base} system with a \SI{1}{\mebi\byte} \gls{llc}, our 256-window coalescer even shows \SI{2}{\%} less in off-chip traffic on average.

\subsection{Area And On-chip Efficiency}
\label{subsec:effi}

% \begin{figure}[t!]
%   \centering
%   \includegraphics[width=1\linewidth]{figures/spmv-soa.png}
%   \caption{\gls{spmv} efficiency comparison with SoA vector processors}
%   \label{fig:spmv-soa}
% \end{figure}

\begin{figure}[t!]
\centering
    
  \subfloat[{\ap} adapter breakdown]{
    \label{fig:area_break}
    \includegraphics[width=0.47\linewidth, height=0.321\linewidth]{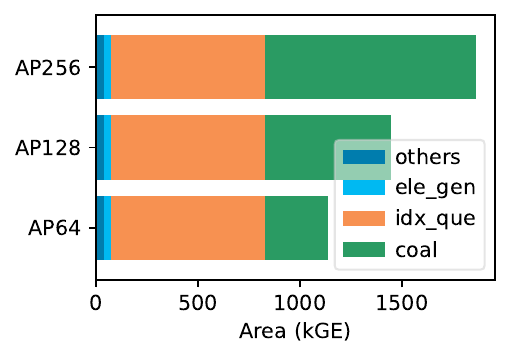}
  }
  \subfloat[\gls{spmv} efficiency comparison]{
    \label{fig:spmv-soa}
    \includegraphics[width=0.47\linewidth, height=0.321\linewidth]{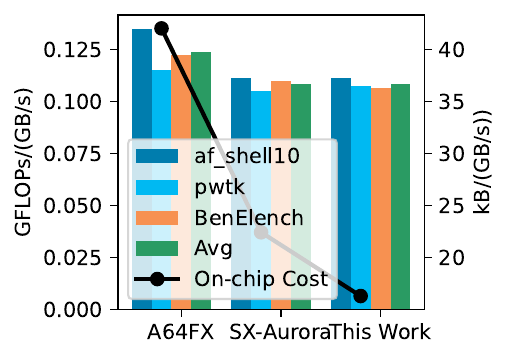}
  }

  \caption{Area And On-chip Efficiency}
  \label{fig:area_soa}
\end{figure}

We implemented our {{\ap}} adapter with Synopsys' \emph{Fusion Compiler} for GlobalFoundries' \SI{12}{\nano\meter} FinFet technology at \SI{1}{\giga\hertz}, worst-case conditions. We mapped the \emph{index queues} of the indirect stream unit and the \emph{hitmap queue} in the coalescer to dual-port \gls{sram} macros, while other queues were mapped to standard cells. The logic area breakdown of the {\ap} adapter is shown in \Cref{fig:area_break}. The index queues take a large portion of the area in the adapter design, up to \SI{754}{\kilo\GE}. The area of the coalescer increases linearly with the coalesce window, resulting in 307, 617, and \SI{1035}{\kilo\GE} for 64, 128, and 256 windows, respectively. We then successfully implemented our 64, 128, 256 coalesce window adapter in \SI{0.19}, \SI{0.26}, \SI{0.34}{\milli\metre\squared} design area with \SI{60.5}, \SI{56.5}, \SI{56.4}{\%} standard cell utilization, respectively. 

Our adapter's minimal on-chip storage of \SI{27}{\kilo\byte} and compact design highlights its potential for on-chip efficient general-purpose processors handling \gls{spmv}. Using a moderately-sized \gls{l2} \gls{spm} and the {\ap} indirect stream unit, our vector processor approach proves more on-chip efficient than extensive and dedicated multi-level memory system hierarchies. \Cref{fig:spmv-soa} compares our method with leading \gls{hbm}-based vector processors, \emph{SX-Aurora}~\cite{sx-aurora} and \emph{A64FX}~\cite{a64fx}, in terms of on-chip storage per memory bandwidth and \gls{spmv} performance per memory bandwidth. For a fair comparison, the on-chip storage accounts for the entire on-chip memory system, including the register file, L1, L2 caches, and \gls{llc}. Memory bandwidth is defined as the maximum achievable main memory bandwidth, as evaluated by the \emph{STREAM} benchmark's \emph{copy} test~\cite{UVAStream}, while the \gls{spmv} performance is based on the same set of sparse matrices in \gls{sell} and \gls{sell}-variant formats. Results for \emph{SX-Aurora} and \emph{A64FX} are taken from the respective references. Our solution offers up to 1.4~\x and 2.6~\x better on-chip efficiency and retains 1~\x and 0.9~\x \gls{spmv} performance efficiency when compared to \emph{SX-Aurora} and \emph{A64FX}, respectively.

%
% Related Work
%
\section{Related Work}
\label{sec:relwork}

\subsubsection{Hardware for Memory Coalescing}
% Memory coalescing is a pivotal mechanism that optimizes memory accesses by consolidating multiple requests into fewer, broader memory transactions. In GPUs, a dedicated coalescer unit close to the L1 cache is employed to manage multiple thread accesses within a warp, enhancing bandwidth utilization and performance\cite{gpuCoal}. Caches also inherently act as hardware coalescers in terms of reducing redundant memory accesses, by capitalizing on cache line reuse. Moreover, dynamic memory coalescing for hybrid memory cube (HMC) has been proposed to coalesce requests to HMC accesses with dynamic granularity\cite{dmc}. Unlike these approaches, our work combines hardware coalescer with near-memory indirect streams. This combination eliminates the overhead of core-side index processing, while coalescing memory requests in an on-chip efficient way compared to large cache hierarchies. We also explore the parallel coalescing process in contrast to the sequential coalescence in the dynamic memory coalescing\cite{dmc}.

Memory coalescing, a mechanism that optimizes memory accesses by consolidating multiple requests into fewer, broader memory transactions, has seen various implementations. In GPUs, a coalescer near the L1 cache manages warp accesses, enhancing bandwidth efficiency\cite{gpuCoal}. Caches also serve as inherent hardware coalescers by reusing cache lines. A distinct approach involves dynamic memory coalescing for hybrid memory cube (HMC) accesses \cite{dmc}, which are processed sequentially. Unlike these, our method integrates a hardware coalescer with near-memory indirect streams, reducing core-side index processing overhead and offering a more area-efficient parallel coalescing scheme.
% and the sequential approach seen in dynamic memory coalescing \cite{dmc}.
% We also introduce parallel coalescing as opposed to the sequential approach seen in dynamic memory coalescing \cite{dmc}.

\subsubsection{Near-memory Extensions}
To accelerate indirect access for \gls{spmv}, many studies have proposed handling it near the main memory with specific hardware units~\cite{axi-pack,Carter1999ImpulseBA,Lloyd2015InMemoryDR,li2023accelerating,scu,tdgraph,menda}. Apart from works relied on narrow channels, as mentioned in \Cref{sec:intro}, some recent works have shifted to broader interfaces. 
% However, these studies often rely on narrow channels with low access granularity, smaller than \SI{64}{\bit}, that are commonly provided by individual chips in older \gls{dram} standards. Carter, John, et al. proposed a smart memory controller called \emph{Impulse} that gathers indirect access at the memory controller to form a dense cache line\cite{Carter1999ImpulseBA}. This is scheduled on multiple \glspl{sdram} with a 64\si{\bit} granularity. Similar works such as the \emph{gather-scatter memory system}~\cite{tanabe2011memory}, gather sparse elements from the narrow channels of single \gls{ddr3} chips. However, these solutions lack portability and flexibility when it comes to modern and standard \gls{dram} interfaces.
% Recently, some studies have moved away from depending on narrow channels~\cite{scu,tdgraph,menda}. 
For example, the \emph{Stream Compact Unit (SCU)}~\cite{scu} compacts sparse vector elements into an array in consecutive memory using an interface with cache-line granularity of \SI{512}{\bit}. While they also explore the coalescence of vector elements, these methods only perform sequential coalescing within a 32-element window, achieving only \SIrange{20}{40}{\%} memory bandwidth utilization~\cite{scu}. Similar works, like \emph{TDGraph}~\cite{tdgraph} and \emph{MeNDA}~\cite{menda}, both employ coalescence when fetching sparse elements from a wide interface to \gls{l2} memory and \gls{dimm} Rank, respectively. However, they too rely on sequential coalescing within a limited window of fewer than 40 elements~\cite{tdgraph,menda}. To the best of our knowledge, this work is the first to leverage both parallel indexing and parallel coalescing for accelerating indirect access near main memory and to delve deeply into their effects.

%
% Conclusion
%
\section{Conclusion}
We proposed a low-overhead coalescer aid for AXI-Pack to efficiently stream indirect accesses from a high-granularity DRAM interface. By leveraging memory-level parallelism and coalescence, it improves indirect stream bandwidth by 8x and offers an end-to-end speedup for SPMV by 3x. Remarkably, its solution for SPMV also demonstrates 1.4x and 2.6x on-chip efficiency compared to SoA vector processors, while retaining 1x and 0.9x SPMV efficiency.
% \todo{\lipsum[1-2]}

%
% Acknowledgment
%
% \section*{Acknowledgment}
% \label{sec:ack}

% \ifreviewmode
%     \textit{Acknowledgments omitted for blind review.}
% \else
%     %Funding
%     \todo{This work was supported in part through funding from the %
%     European High Performance Computing Joint Undertaking (JU) %
%     under Framework Partnership Agreement No 800928 and %
%     Specific Grant Agreement No: 101036168 (EPI SGA2) and No: 101034126 (The EU Pilot).} %
% \fi

\bibliographystyle{IEEEtran}
\bibliography{IEEEabrv,main}

\end{document}

%% file: acronyms.tex
\newacronym{csr}{CSR}{compressed sparse row}
\newacronym{sell}{SELL}{sliced ELLPACK}
\newacronym{hbm}{HBM}{high bandwidth memory}
\newacronym{lpddr}{LPDDR}{low-power double data rate}
\newacronym{spmv}{SpMV}{sparse matrix vector multiplication}
\newacronym{cpu}{CPU}{central processing unit}
\newacronym{gpu}{GPU}{graphics processing unit}
\newacronym{axi}{AXI4}{Advanced eXtensible Interface 4}
\newacronym{llc}{LLC}{last-level cache}
\newacronym{sram}{SRAM}{static random-access memory}
\newacronym{dram}{DRAM}{dynamic random-access memory}
\newacronym{sdram}{SDRAM}{synchronous dynamic random-access memory}
\newacronym{cshr}{CSHR}{coalescer status holding register}
\newacronym{vpc}{VPC}{vector processor core}
\newacronym{vps}{VPS}{vector processor system}
\newacronym{l2}{L2}{level two}
\newacronym{spm}{SPM}{scratch pad memory}
\newacronym{vmac}{VMAC}{vector multiply add}
\newacronym{rtl}{RTL}{register transfer level}
\newacronym{ddr3}{DDR3}{double data rate 3}
\newacronym{dimm}{DIMM}{dual inline memory module}
\newacronym{fifo}{FIFO}{first in, first out}
\newacronym{rvv}{RVV}{{RISC-V} vector extension}